\documentclass[12pt]{article}
\usepackage{graphicx}
\usepackage{amsfonts,amsthm}
\usepackage{amsmath,amssymb}
\usepackage{bm}

\newcommand{\be}{\begin{equation}}
\newcommand{\ee}{\end{equation}}
\newcommand{\bea}{\begin{eqnarray}}
\newcommand{\eea}{\end{eqnarray}}

\newcommand{\bk}{{\bf k}}
\newcommand{\bK}{{\bf K}}

\newcommand{\bp}{{\bf p}}

\newcommand{\br}{{\bf r}}

\def\lsim{\mathrel{\rlap{\lower4pt\hbox{\hskip1pt$\sim$}}
		\raise1pt\hbox{$<$}}}         %less than or approx. symbol
\def\gsim{\mathrel{\rlap{\lower4pt\hbox{\hskip1pt$\sim$}}
		\raise1pt\hbox{$>$}}}         %greater than or approx. symbol

\title{Double-twisted spectroscopy with delocalized atoms}

\author{Igor~P.~Ivanov\thanks{E-mail: ivanov@mail.sysu.edu.cn}\\
	{\small School of Physics and Astronomy, Sun Yat-sen University, 519082 Zhuhai, China}
}

\begin{document}
	
	\maketitle
	
\begin{abstract}
Interaction of atoms with twisted light is the subject of intense experimental and theoretical investigation.
In almost all studies, the atom is viewed as a localized probe of the twisted light field.
However, as argued in this paper, conceptually novel effects will arise if light-atom interaction is studied
in the double-twisted regime with delocalized atoms, that is,
either via twisted light absorption by atom vortex beam,
or via two-twisted-photon spectroscopy of atoms in a non-vortex but delocalized state.
Even for monochromatic twisted photons and for an infinitely narrow line, 
absorption will occur over a finite range of detuning. 
Inside this range, a rapidly varying absorption probability is predicted,
revealing interference fringes induced by two distinct paths leading to the same final state.
The number, location, height and contrast of these fringes can give additional information 
on the excitation process which would not be accessible in usual spectroscopic settings.
Visibility of the predicted effects will be enhanced at the future Gamma factory 
thanks to the large momenta of ions.
\end{abstract}

\section{Light-atom interaction in double-twisted regime}

\subsection{Introduction}

Interaction of atoms with twisted light is a fascinating, intensely studied chapter in atomic physics.
Twisted light refers to solutions of Maxwell's equations exhibiting an optical vortex,
the light field with azimuthally dependent phase around a line of phase singularity.
At quantum level, it is described by twisted, or vortex, photons carrying
non-zero orbital angular momentum (OAM) with respect to their propagation direction.
Twisted light is known since decades \cite{Allen:1992zz,Paggett:2017,Knyazev:2019} 
and has already found numerous applications \cite{applications,andrews2012angular}.
Other forms of structured light beyond optical vortices \cite{rubinsztein2016roadmap,forbes2021structured} are also being explored.

\medskip

If an atom is placed in the light field of an optical vortex, its behavior as a whole
as well as its internal transitions display novel effects, 
see the recent review \cite{babiker2018atoms} and references therein.
For example, if placed close to the phase singularity line, the atom is subject to a very strong phase gradient.
When absorbing a photon, it can get a ``superkick'', acquiring a transverse momentum larger 
than the transverse momentum of any plane wave component of the twisted light beam \cite{barnett2013superkick,Afanasev2021superkick}.
Also, a photon with non-zero OAM absorbed by the atom can excite multipole transitions 
which are suppressed for plane wave photons.
Theoretical investigation of this process began almost two decades ago \cite{babiker2002orbital},
leading to its experimental demonstration in \cite{schmiegelow2016transfer}.
Position-dependent modifications of the atomic transition selection rules 
were further investigated in \cite{afanasev2018experimental,solyanik2019excitation,schulz2019modification,schulz2020generalized}.
Ionization of atoms by twisted photons \cite{Matula:2013,kaneyasu2017limitations},
including two-photon ionization with a combination of twisted and plane-wave photons \cite{Kosheleva:2020},
as well as scattering of twisted light by atoms \cite{Peshkov:2018}
were also theoretically explored.

\medskip 

In all these studies, the atom is viewed as a localized, almost pointlike probe 
of the optical vortex light field. This is a valid assumption for most experimental settings
with trapped atoms or atomic beams.
However, one can also consider atoms prepared in {\em delocalized} states of the spatial extent at least of the order of optical wave length. 
In particular, one can consider {\em atom vortex beams}, in which the center of mass wave function
exhibits a phase vortex and carries an OAM.
Although the idea was discussed long ago \cite{helseth2004atomic}, 
the exact Bessel or Laguerre-Gaussian atomic vortex states
were scrutinized only in the last decade \cite{hayrapetyan2013bessel,lembessis2014atom}, 
prompted by the successful demonstration of electron vortex beams \cite{Uchida:2010,Verbeeck:2010,McMorran:2011}.
It was suggested that vortex atom beams could be produced by diffraction of 
a non-vortex delocalized atom wave through a light mask 
formed by the interference of a Laguerre-Gaussian beam with a plane-wave or Gaussian beam.
The state-of-the-art in this field was summarized in section~14 of the review \cite{babiker2018atoms}.

\medskip

Very recently, vortex beams of individual atoms were finally demonstrated experimentally \cite{luski2021vortex}
with the aid of an array of conventional fork diffraction gratings of submicron size, 
not via light-atom interaction.
Thus, experiments probing atom vortex beams with light is still an uncharted territory.

\subsection{The proposal}

In all situations outlined above, 
we deal with only one initial particle in a vortex state: it is either twisted light probed by localized atoms 
or an atom vortex beam interacting with light beam.
This setting can be called the single-twisted regime.
In this paper I advocate exploring light-atom interaction in the {\em double-twisted} regime,
which can be achieved as
\begin{itemize}
	\item 
	twisted light interacting with vortex atom beam (twisted atoms), 
\begin{equation}
\mbox{atom}_{tw} + \gamma_{tw} \to \mbox{atom}^*,\label{process1}
\end{equation}
	\item
	or two-twisted-photon spectroscopy of non-vortex but delocalized atom beams
\begin{equation}
\mbox{atom} + \gamma_{tw} + \gamma_{tw} \to \mbox{atom}^*.\label{process2}
\end{equation}
\end{itemize}
Here, by labeling the final state just as an excited atom, I deliberately omit the subsequent evolution of the final state.
This is done to highlight the fact that the effects to be discussed arise due to non-trivial interference 
\textit{at the moment of excitation} and will persist for any outcome of the photon absorption. 
Thus, I propose not one specific experiment but a class of atom-light interaction processes with different final states.
The key features that unites them all are, first, the presence of two twisted initial states and, second, the necessity of working with 
delocalized atomic states, not pointlike atoms.

\medskip

Experimental realization of such collision setting can be rather challenging.
The main motivation for striving for such collisions is the physics opportunities they bring.
As I will outline in this paper, one can explore a kinematic regime which cannot be reproduced
with plane wave or Gaussian beams, nor ins single-twisted configurations.

\subsection{The benefits of double-twisted regime}\label{section-benefits}

Any wave function can be represented as a superposition of plane waves.
The phase vortex in a twisted state --- be it a photon or an atom vortex beam --- arises
from delicate interference among infinite amount of plane wave components.
However if only one particle is twisted and all other particles or fields 
are approximated by plane waves, each plane wave component leads to 
a kinematically distinct final state. Different final states do not interfere, 
and the delicate interference pattern of the initial state is lost, or at best hidden.

\medskip

Collisions of \textit{two} initial particles in twisted states lead to novel effects,
which so far have been theoretically explored only in particle physics realm.
They can be purely kinematical \cite{Ivanov:2011kk,Ivanov:2016oue,Karlovets:2016jrd,Ivanov:2019pdt}
or can involve spin degrees of freedom in a novel way \cite{Ivanov:2019vxe}. 
The origin of these effects can be tracked to the famous Young's two slit experiment 
but now seen in momentum space \cite{Ivanov:2016jzt}.
Since each of the two twisted states is a superposition of many plane waves,
one can reach the same final state via two distinct ``paths'' in momentum space.
As a result, one can observe interference effects between two plane-wave scattering amplitudes
originating from different initial momenta and leading to the same final state. 
These effects can be observed as interference fringes in various kinematical dependences, 
which could not be seen in the plane-wave or single-twisted regimes.

\medskip

In the present paper, I propose to explore the atomic physics counterparts of these effects.
Unlike particle physics, where twisted high-energy particles are still far from experimental realization,
the double-twisted regime in atomic physics can be achieved with existing technology or in near future.
In particular, the proposed Gamma factory at CERN \cite{GammaF-for-atomic} can be advantageous
in realizing two-twisted-photon spectroscopy.

\section{Absorbing two twisted photons}

\subsection{Assumptions taken}

Although the two versions \eqref{process1} and \eqref{process2} of the double-twisted regime
involve different atomic transitions, they share similar kinematical effects.
To illustrate the main idea, let us consider the two-twisted-photon spectroscopy \eqref{process2},
just because it does not rely on twisting the atomic beams and, therefore, may be realizable 
with existing technology. To simplify the analysis, 
let us assume that the two photons are in Bessel states and neglect their polarization degrees of freedom.
The initial atom will be described as a freely propagating plane wave. 
As mentioned above, this unconventional assumption is taken not because it is easily realizable in atomic physics experiments
but because it will help demonstrate the effects proposed here as clearly as possible.
%Thus, delocalized atoms is a goal to strive for, not a routine starting point.
Also, most calculations will be done for non-relativistic atoms;
extension to the fully relativistic case is straightforward. 

\medskip

As already mentioned above, we consider the final state as an excited atom and do not track
the actual outcome of the photon absorption.
Any specific outcome --- be it scattering, ionization, or fluorescence --- will introduce more
kinematical variables to explore and will offer even {\em richer} physics opportunities. 
What I stress is that even in the most restricted version, with the total final state momentum as the only available variable,
we can observe effects which could not be achieved in fully plane wave or twisted + plane wave cases.

\medskip

With all these reservations, the process of two twisted photon absorption can be 
considered as a $3 \to 1$ collision.
The initial state contains an atom in the ground state with the (invariant) mass $M_i$, three-momentum $\bp_i$,
and the energy $E_i \approx M_i + \bp_i^2/2M_i$.
It also contains two monochromatic twisted photons defined with respect to a common axis $z$.
Their energies are $\omega_1$ and $\omega_2$,
longitudinal momenta are $k_{1z} > 0$ and $k_{2z} < 0$, the moduli of their 
transverse momenta are $\varkappa_1$ and $\varkappa_2$, and the OAM $z$-projections are $m_1$ and $m_2$.
The final atom has the mass $M_f = M_i + E_{exc}$, the three-momentum $\bp_f$,
and the energy $E_f \approx M_f + \bp_f^2/2M_f$.
For simplicity, the two-photon energy level is assumed to be infinitely narrow, 
so that the excitation energy $E_{exc} = M_f - M_i$ is fixed.
A finite line width --- and novel ways to scan it without changing photon energies ---  will be mentioned below.

\subsection{The plane-wave case}

As a reference, let us begin with the usual fully plane-wave (PW) case, in which the two
photons have definite 3-momenta $\bk_1$ and $\bk_2$. 
The scattering matrix element can be written generically as
\begin{equation}
S_{PW} \propto \delta(E_i + \omega_1 +\omega_2 - E_f)\delta^{(3)}(\bk_1 + \bk_2 - \bK)\cdot {\cal M}\,,\label{S-PW} 
\end{equation}
with $\bK \equiv \bp_f - \bp_i$ being the momentum transfer.
The transition amplitude ${\cal M}$ incorporates all the internal dynamics of the excitation process
and can depend on the kinematic parameters just introduced. 
However, its exact form will be unimportant.
Squaring the scattering amplitude and regularizing the squares of delta-functions in the usual way \cite{LL4}, 
we obtain the following generic expression for the PW cross section: 
\begin{eqnarray}
d\sigma &\propto & \delta(E_i + \omega_1 +\omega_2 - E_f) |{\cal M}|^2  \,  \delta^{(3)}(\bk_1 + \bk_2 - \bK) \, d^3 K\,,\nonumber\\[2mm]
\sigma &\propto & \delta(E_i + \omega_1 +\omega_2 - E_f) |{\cal M}|^2\,.\label{sigma-PW-0}
\end{eqnarray}
Notice that the momentum transfer to the atom is fixed at $\bK = \bk_1 + \bk_2$ and the dependence
on the total energy of the colliding particles is proportional to $\delta(E_i + \omega_1 +\omega_2 - E_f)$.
Since the atom energy depends on its velocity, we get
\begin{equation}
\omega_1 + \omega_2 = E_f - E_i = E_{exc} + \frac{\bK(2\bp_i + \bK)}{2M}\,.\label{E-conservation}
\end{equation}
Here, we replaced $M_i, M_f \to M$ in the denominator since we stick to the non-relativistic approximation.

\medskip

There are two obvious but important conclusions we draw from this well known expression.
First, the excitation occurs only when the sum of laser photon energies exactly matches
the excitation energy offset by the recoil effects. If we scan a certain region, say, in $\omega_1$,
keeping all other parameters fixed, we will observe a delta-function spike of the cross section
at the moment when \eqref{E-conservation} is fulfilled.
Second, if the initial PW photon momenta combine to a non-zero $\bk_1 + \bk_2 = \bK$, 
the two-photon absorption will affect only atoms with a fixed value $\bK \bp_i$.

\subsection{Two Bessel photons}

If the initial photons are in Bessel twisted states,
we represent each photon, neglecting its polarization degrees of freedom, as a superposition of plane waves 
\cite{Jentschura:2010ap,Jentschura:2011ih,Ivanov:2011kk,Karlovets:2012eu}:
\begin{equation}
|\varkappa,m\rangle = e^{-i \omega t + i k_z z} \int {d^2 k_\perp \over(2\pi)^2}a_{\varkappa m}(\bk_\perp) e^{i\bk_\perp \br_\perp}\,,
\label{twisted-def}
\end{equation}
where
\begin{equation}
a_{\varkappa m}(\bk_\perp)= (-i)^m e^{im\varphi_k}\sqrt{2\pi \over \varkappa}\; \delta(|\bk_\perp|-\varkappa)\label{a}
\end{equation}
is the corresponding Fourier amplitude. 
The $S$-matrix element of the two twisted photon excitation is 
\begin{equation}
S = \int {d^2 \bk_{1\perp} \over (2\pi)^2} {d^2 \bk_{2\perp} \over (2\pi)^2} 
a_{\varkappa_1 m_1}(\bk_{1\perp}) a_{\varkappa_2, -m_2}(\bk_{2\perp}) S_{PW}\,.\label{S-tw}
\end{equation}
Substituting here the Fourier amplitudes of the Bessel states, we get, schematically,
\begin{equation}
S \propto \delta(\Sigma E)\cdot \delta(\Sigma k_z) \cdot {\cal J}\,,\label{S-tw2}
\end{equation}
where $\delta(\Sigma E) \equiv \delta(E_i + \omega_1 +\omega_2 - E_f)$, $\delta(\Sigma k_z) \equiv \delta(k_{1z}+k_{2z}-K_z)$
and the twisted amplitude ${\cal J}$ is
\begin{eqnarray}
{\cal J} &=& \int d^2 \bk_{1\perp} d^2 \bk_{2\perp} \, e^{im_1\varphi_1 - im_2\varphi_2}\,
\delta(|\bk_{1\perp}|-\varkappa_1) \delta(|\bk_{2\perp}|-\varkappa_2)
\delta^{(2)}(\bk_{1\perp}+\bk_{2\perp} - \bK_\perp)\cdot {\cal M}\nonumber\\
&=&\varkappa_1\varkappa_2 \int d\varphi_1 d\varphi_2\, e^{im_1\varphi_1 - im_2\varphi_2}\,
\delta^{(2)}(\bk_{1\perp}+\bk_{2\perp} - \bK_\perp)\cdot {\cal M}\,.\label{J}
\end{eqnarray}
\begin{figure}[!ht]
	\centering
	\includegraphics[width=0.7\textwidth]{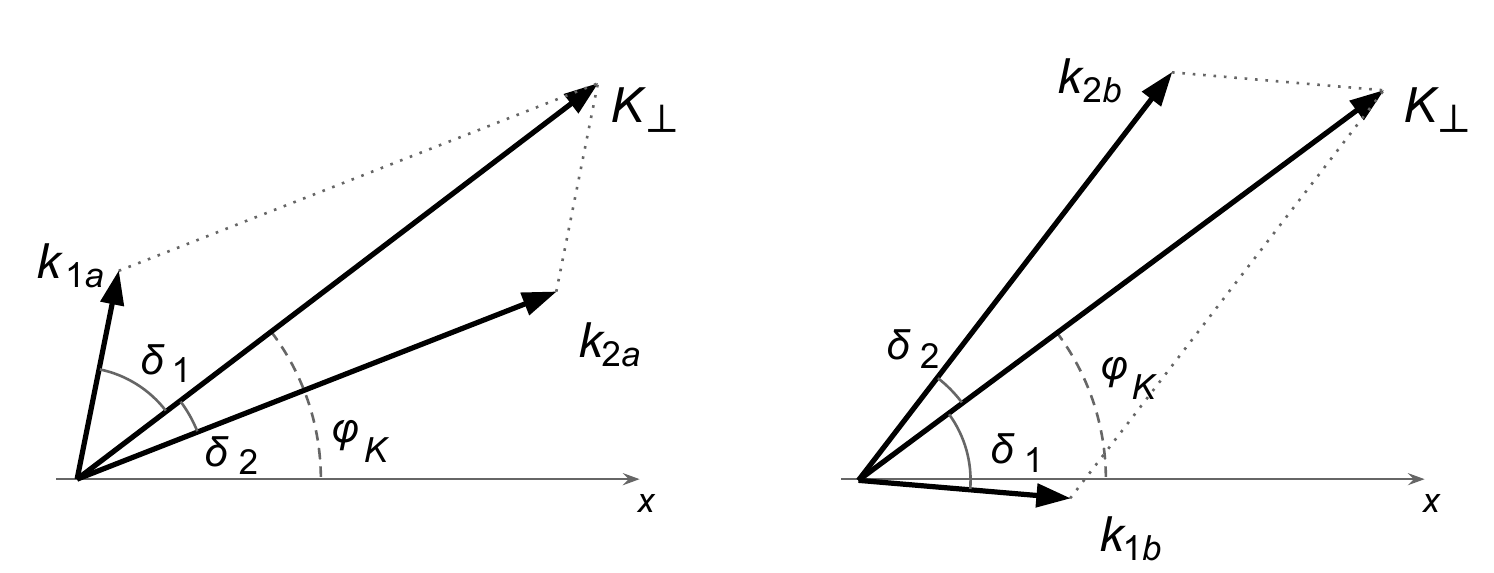}
	{\caption{\label{fig-2configurations} The two kinematic configurations in the transverse plane which satisfy
			the transverse momentum conservation law in collision of two Bessel states.}}
\end{figure}
Thus, only the longitudinal component of the momentum transfer $K_z$ is fixed by the initial state kinematics.
The transverse momentum transfer $\bK_\perp$ displays a distribution.
To see it, we evaluate the integral \eqref{J} \cite{Ivanov:2011kk}: 
it is non-zero if and only if $\varkappa_i = |\bk_{i\perp}|$ 
and $K_\perp \equiv |\bK_\perp|$ satisfy the triangle inequalities
\begin{equation}
|\varkappa_1 - \varkappa_2| \le K_\perp \le \varkappa_1 + \varkappa_2\,.\label{ring}
\end{equation}
and form a triangle with the area $\Delta = (2 K_\perp^2\varkappa_1^2 + 2 K_\perp^2\varkappa_2^2 + 2\varkappa_1^2\varkappa_2^2 
- K_\perp^4 - \varkappa_1^4 - \varkappa_2^4)^{1/2}/4$.	
Out of many PW components in each twisted photon, 
the integral \eqref{J} receives contributions only from two PW combinations shown in Figure~\ref{fig-2configurations}
with the following azimuthal angles:
\begin{eqnarray}
\mbox{configuration a:} &&\varphi_1 = \varphi_{K} + \delta_1\,,\quad \varphi_2 = \varphi_{K} - \delta_2\,,\nonumber\\
\mbox{configuration b:} &&\varphi_1 = \varphi_{K} - \delta_1\,,\quad \varphi_2 = \varphi_{K} + \delta_2\,.\label{phi12}
\end{eqnarray}
Notice that $\delta_1$ and $\delta_2$ are not azimuthal variables but
the internal angles of the triangle with the sides $\varkappa_1$, $\varkappa_2$, $K_\perp$. 
The \textit{exact} result for the twisted amplitude ${\cal J}$ can be compactly written as
\begin{equation}
{\cal J} = e^{i(m_1 - m_2)\varphi_{K}}{\varkappa_1 \varkappa_2 \over 2\Delta}
\left[{\cal M}_{a}\, e^{i (m_1 \delta_1 + m_2 \delta_2)} + {\cal M}_{b}\, e^{-i (m_1 \delta_1 + m_2 \delta_2)}\right].\label{J2}
\end{equation}
Notice that the plane-wave amplitudes ${\cal M}_{a}$ and ${\cal M}_{b}$ are calculated for the two distinct initial momentum 
configurations shown in Figure~\ref{fig-2configurations} but for the same momentum transfer $\bK$.
Since they lead to the same final state, the two PW configurations interfere.
Squaring \eqref{S-tw} and following the standard regularization 
\cite{Jentschura:2010ap,Jentschura:2011ih,Ivanov:2011kk,Karlovets:2012eu},
we obtain the generalized cross section in the form
\begin{equation}
d\sigma \propto \delta(\Sigma E)\cdot \delta(\Sigma k_z)\, |{\cal J}|^2 \, d^3K \ 
=\  \delta(E_i + \omega_1 +\omega_2 - E_f)\, |{\cal J}|^2 \, d^2\bK_\perp\,.\label{dsigma-tw}
\end{equation}
The energy conservation law expressed in Equation~\eqref{E-conservation} still holds.
However, unlike the plane-wave collision where the final momentum was completely fixed 
by the initial kinematics, we have here a \textit{distribution} over $\bK_\perp$
inside the annular region defined by Equation~\eqref{ring}. 
The energy conservation only fixes the value of
\begin{equation}
\bK^2 + 2\bp_i\bK = K_\perp^2 + K_z^2 + 2p_{iz}K_z + 2\bp_{i\perp}\bK_\perp = 2M \delta \,,\label{fixed-quantity}
\end{equation}
where $\delta = \omega_1+\omega_2-E_{exc} $ is the detuning of the two-photon transition.
Thus, the value of the detuning $\delta$ is not fixed but depends on $\bK_\perp$.

\section{Physics insights}
\subsection{Where does the excited atom go?}

Let us now understand the implications of the condition~\eqref{fixed-quantity}.
Suppose the momentum of the atom is very small, $p_i \ll \varkappa$.
Then the transition occurs when the momentum transfer 
\begin{equation}
\bK^2 = K_\perp^2 + K_z^2 = 2 M\delta\,.\label{fixed-quantity-2}
\end{equation}
When absorbing two photons, the atom gets a kick.
Since $K_z = k_{1z} + k_{2z}$ is fixed, this relation fixes the polar angle $\theta_K$ of the excited atom
$\cos\theta_K = K_z/\sqrt{2M\delta}$.
In the simplest case when the amplitudes ${\cal M}_a = {\cal M}_b = {\cal M}_0$, 
the cross section reduces to 
\begin{equation}
d\sigma \propto {|{\cal M}_0|^2 \over \Delta^2} \cos^2(m_1 \delta_1 + m_2 \delta_2) \,d\varphi_K\label{dsigma-tw3}
\end{equation}
with a uniform $\varphi_K$ distribution.
If ${\cal M}_a \not = {\cal M}_b$, which may be the case when the photon polarization is taken into account,
additional non-trivial effects can be expected.

\medskip

Here comes the key feature of this process. Although we work with monochromatic photons and atoms
and an infinitely narrow line, excitation takes place within a \textit{certain range} of the initial energies.
In terms of detuning $\delta$, we get the interval
\begin{equation}
(\varkappa_1 - \varkappa_2)^2 + K_z^2 \le 2 M \delta \le (\varkappa_1 + \varkappa_2)^2 + K_z^2\,,\label{E-range}
\end{equation}
inside which the cross section rapidly oscillates as $\cos^2(m_1 \delta_1 + m_2 \delta_2)$. 
These are the interference fringes of the total excitation rate mentioned above. They can be observed either as the variation 
of the excitation rate during an energy scan or as a set of similar interference fringes 
in the polar angle $\theta_K$ distribution across a broad range of its values.
The number, height, location and contrast of these fringes depend on the OAM values $m_i$, 
on the absolute values of the transverse momenta $\varkappa_i$, and, for ${\cal M}_a \not = {\cal M}_b$, 
on the relative phase between these two amplitudes. These are the atomic physics counterparts of the fringes 
derived for high-energy particle collisions in \cite{Ivanov:2019pdt}.

\subsection{Fourier analysis of the lineshape}

The interference fringes just mentioned are produced by a {\em single}, infinitely narrow two-photon transition line. 
They arise due to production of the same final state by initial photon pairs in two different PW configurations,
the two ``paths'' mentioned in section~\ref{section-benefits}. 
These fringes cannot arise in the usual PW case nor in the PW$+$twisted configuration
considered in \cite{Kosheleva:2020}. 
This is not a quantitative deviation from the PW case but a qualitatively new feature, 
a new dimension in the final state to explore.

\medskip

These fringes give rise to an unconventional spectroscopic tool which could be called the Fourier analysis of a lineshape.
Indeed, if an infinitely narrow line produces a $\cos^2(m_1 \delta_1 + m_2 \delta_2)$ oscillation pattern
within the detuning window, then a line of non-zero width and a non-trivial profile
will produce a superposition of such oscillation patterns with different shifts.
If the resulting oscillation pattern is measured with sufficient precision  
and if the parameters of the twisted light beams $m_i$ and $\varkappa_i$ are known precisely, 
one can solve the inverse problem and reconstruct the line shape.
We stress that this information arises from a series of experiments with different $m_i$ 
at \textit{fixed photon energies}. That is, line shape can be recovered without the need for energy scan.

\subsection{The role of high initial momentum}

To estimate the width of the detuning window inside which one could observe interference fringes,
let us consider tightly focused Bessel photons with $\varkappa_i,K_{iz} \sim {\cal}O(0.1\ \mbox{eV})$.
For light atoms, $M \sim 10$ GeV, it gives typical detuning range of the order of $\delta \sim 10^{-12}$ eV.
All the interference fringes are squeezed inside this window.
Although they become well visible in the $\theta_K$ angular distribution, such an experiment would still require
all energies to be fixed with accuracy better than $10^{-12}$ eV, which is challenging.

\medskip

The observability improves dramatically if the initial atom carries a significant momentum $\bp_i \gg \varkappa_i$.
Since atoms are heavy, this condition is satisfied even for non-relativistic atoms.
Within the crossed beam scenario, the atom moves perpendicularly to the photon axis with momentum $p_i=|\bp_{i\perp}| = \beta M$.
If the two twisted photons have balanced longitudinal momenta $k_{1z}+k_{2z} =0$,
then the window in $\delta$ grows to
\begin{equation}
-(\varkappa_1+\varkappa_2)\beta \le \delta \le (\varkappa_1+\varkappa_2)\beta\,. \label{range2}
\end{equation}
For mildly relativistic atoms, this range is of the order of $0.1$ eV.

\medskip

If there is a transition line anywhere within this window, it will lead to a non-zero transition rate.
If there are two or more transition lines residing within this window, 
they will be simultaneously excited even if the experiment runs at fixed photon energies.
If we wish to find the exact location or study the shape of this lines, we could repeat the experiment with different 
values of $m_i$ keeping the energy fixed. Of course one could also vary the laser wavelength 
and measure how the excitation rate depends on the photon energies; such a scan would produce a fringe-like
oscillation as a function of photon energies.

\medskip

In the realm of relativistic ions offered by the future Gamma factory, 
the effects discussed here will stay intact. A fully relativistic counterpart of the above calculation 
shows that the region of $\delta$ can be as large as Equation~\eqref{range2} with $\beta\to 1$
in the crossed beam scenario.
An accurate realistic calculations of transition amplitudes are needed in order to make quantitative predictions.

\section{Conclusions}

To summarize, I proposed here to explore light-atom interaction with delocalized atom beams in the double-twisted configuration.
This can be done either by shining twisted light on an atom vortex beam of comparable transverse size,
or by exciting a two photon transition of a delocalized non-vortex atoms with two aligned twisted photons.
Realizing such experiment would be a challenge, but there is physics pay-off: 
both setting lead to a remarkable, yet unexplored kind of interference,
which may offer new insights into light-atom interactions.
The visibility of the effects will be boosted if relativistic atomic beams are used. 
The future Gamma factory is a promising site for such experiments, but 
the proof-of-principle experiments can be done even with lower energy atoms or ions.

\medskip

Using delocalized atoms, either as an atom vortex beam or as a non-vortex Gaussian beam 
with a sufficiently large transverse coherence length, is a challenging requirement.
However, the very recent success in producing helium vortex beams inspires optimism \cite{luski2021vortex}.
Assuming that vortex beams of other atoms could be produced, either in a similar fashion
or using the long-anticipated atom-light interaction \cite{hayrapetyan2013bessel,lembessis2014atom},
one can perform detailed calculations for various atomic processes in double-twisted regime
and pinpoint specific observables most suitable for detecting the novel interference effects.
Such a calculation should also take into account the unavoidable momentum spread 
of delocalized atoms, which will lead to certain smearing of the interference fringes.
It remains to be seen which level of smearing is tolerable for observation of the effects.

\medskip

\textbf{Acknowledgments} \par 
I am thankful to Andrey Surzhykov for his suggestion to take into account the large momentum
of the initial atoms or ions.

\end{document}